\documentclass[conference]{IEEEtran}
\IEEEoverridecommandlockouts
\usepackage{color}
\usepackage{graphicx}  
\usepackage{psfrag}    
\usepackage{algorithm}
\usepackage{algorithmic}
\usepackage{amsfonts,amssymb}
\usepackage{amsthm}
\usepackage{amstext}
\usepackage{bm}
\usepackage{subfigure}
\usepackage{amsmath}
\usepackage{amssymb}
\usepackage{graphicx,cite,epsfig,amssymb,amsmath,subfigure,url,stfloats,latexsym,bm,booktabs,diagbox,clrscode}
\usepackage{algorithm}
\usepackage{algorithmic}
\usepackage{float}

\usepackage{algorithm}
\usepackage{algorithmic}
\usepackage{multirow}
\usepackage{amsfonts}

\usepackage{caption}
\usepackage{graphicx}
\usepackage{epstopdf}
\usepackage{amsfonts}
\usepackage{subfigure}
\usepackage{forest}
\usepackage{tikz-qtree}
\usepackage{tikz-cd}
\usepackage{mathtools}
\usepackage{tikz}
\usepackage{changes}
\definechangesauthor[color=orange]{Yuming}

\usetikzlibrary{arrows,decorations.markings}
\usetikzlibrary{decorations.pathreplacing,angles,quotes,shapes}

\begin{document}
\title{Markovian Analysis of Information Cascades with Fake Agents}
\author{\IEEEauthorblockN{Yuming Han}
\IEEEauthorblockA{Northwestern University, Evanston, USA}
\IEEEauthorblockA{Email: yuminghan2021@u.northwestern.cedu}

}
\maketitle

\begin{abstract}
People often learn from other’s actions when they make decisions while doing online shopping. This kind of observational learning may lead to information cascades, which means agents might ignore their own signals and follow the "trend" created collectively by the actions of their predecessors. It is well-known that with rational agents, such a cascade model can result in either correct or incorrect cascades. In this paper, we additionally consider the presence of fake agents who always take fixed actions and we investigate their influence on the outcome of these cascades. We propose an infinite Markov Chain sequence structure and a tree structure to analyze how the fraction and the type of such fake agents impacts behavior of the upcoming agents. We show that an increase in the fraction of fake agents may reduce the chances of their preferred outcome, and also there is a certain lower bound for the probability of a wrong cascade. In particular, we discuss the probability of an agent being fake tends to 1 and the effect of a constant portion of fake agents.
\end{abstract}


\section{Introduction}

In this paper, we analyze a recommendation-based market where agents sequentially arrive and decide whether to buy a new product that is up for sale based on their own prior knowledge of its quality and by observing the decisions of their predecessors.  Every agent’s decision is recorded as a reference for later agents to a common database so that later agents are able to obtain information. This kind of model was proposed in\cite{label1}-\cite{label3}as a Bayesian learning model where  a  sequence of agents take optimal actions. In this model, every agent comes with his own opinion or signal about this new product followed by a known probbibilty distribution. Besides, there is a given common database contains all previous desicions made by predecessors as a reference. Each newly arrived agent makes a one-time Bayesian rational decision based on his own signal and these observations from the common database, and then his action is recorded.

We emphasize on two main results for such Bayesian model. First, there is learning. It comes from agents make an optimal decision by observing the database to achieve the true value of the product. Second, there is information cascade. Information cascade is an agent ignores his own knowledge and follows the decisions made by his predecessors. The cascade situation would also prevent the upcoming agents from learning from subsequent observations, i.e., once a wrong cascade (socially sub-optimal) occurred, this market would be trapped in the wrong cascade forever. \cite{label1}-\cite{label3}assumes agents have discrete known distribution private signals, which leads to a cascade happens in a finite number of agents with probability one. That gives us a great opportunity to study on the probability of wrong casesde with different model settings. 

In this paper, we consider a similar model as in \cite{label1}-\cite{label3}. In addition to rational agents, we introduce a certain type of agents named "fake agents" who always make a fixed action. Regardless of their own opinion or the actions of prior agents, fake agents will only buy (Y-type) or not buy (N-type) a product to influence the probability of the outcome of a cascade. For example, when people go to Amazon to buy an item, even if this item has a poor quality, a Y-type fake agent would continually buy it and record that this item has good quality on the website (five stars). This information will have an impact for the future agents. On the other hand a N-type fake agent would act in the opposite way to always give a bad review (one star). We study on the influences of these fake agents on the probabilites of wrong information cascades and some certain thresholds appeared in the simulation resutls. An infinite Markov chain model, similar to \cite{label4}, is proposed. In our model, we mainly concentrate on the impact of varying the portion of fake agents on the probability of correct and incorrect cascades. Our work is based on \cite{label5}, Poojary and Berry discuss the scenario of one type of fake agents in the network.

There are some other related work similar to our model, such as \cite{label6}, where the agent's observations depend on an underlying network and \cite{label7}, where the crazy and stubborn agents are invloved. The work of Smith and Sorensen \cite{label8} relaxed both the assumptions of binary signals and homogeneous agents. Another aspect of related work is that which considers different model structure. In \cite{label10} a non-Bayesian model was considered, and an information structure with selected cost was introduced in \cite{label11}.

In Section II, we propose our model. We analyze our model and discuss the cascade results in Section III. Next, in Section IV, we introduce the Markov Chain process. We talk about some certain thresholds, approximations and bounds in Section V. We conclude in Section VI with some directions for future work on modeling and analyzing for realistic settings.

\section{System Model}
We consider a model, similar to \cite{label1}, in which with a countable sequence of agents, indexed $i=1,2,\ldots$ where the index represents both the time and the order of arrival actions for each agent. Every agent i chooses an action $A_{i}$ of either buying (Y) or not buying (N) one item which has a true value (V) that could be either good (G) or bad (B). We consider $V=G$ and $V=B$ to be equiprobable. The prior distribution of the item value is assumed to be common knowledge and the realized value of the item is assumed to be the same for agents. 

Further, each agent $i$ receives a private signal $S_{i}\in \{H(High),L(Low)\} $ that represents its prior knowledge of the product's quality. This private signal $S_{i}$ partially reveals the true value $V$ through a binary symmetric channel (BSC) with crossover probability $1-p$, depicted in Figure 1. Here, $p\in\{0,1\}$ denotes the private signal quality. Agent $i$ then takes a rational action $A_{i}$ based on his private signal $S_{i}$ and the past observations $\{{O_{1},O_{2},\ldots,O_{i-1}}\} $ of actions $\{{A_{1},A_{2},\ldots,A_{i-1}}\} $. Next, we modify the model in \cite{label1} by considering that at each time instant, an agent could be a Y -type fake agent (w.p.) $\epsilon \in [0, 1)$ or a N -type fake agent (w.p.) $\beta \in [0, 1)$ or an ordinary agent w.p. $1-\gamma$. Here, $\gamma =\epsilon + \beta$  denotes the total probability of an agent being fake. An ordinary agent is honest in reporting its actions, i.e., $O_{i} = A_{i}$. Whereas, a Y-type (N-type) fake agent always reports a Y (N).

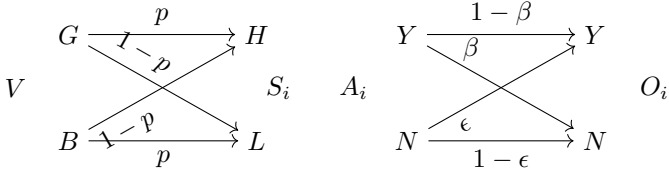
\begin{figure}[h]
\begin{tikzpicture}[scale=0.3]
\tikzstyle{arrow}     = [thick,->,>=stealth]
\node (V)  {$V$};
\node (G)  [above right of=V,node distance=10mm ] {$G$};
\node (B)  [below right of=V, node distance=10mm] {$B$};
\node (H)  [right of=G, node distance=2.5cm] {$H$};
\node (L)  [right of=B, node distance=2.5cm]{$L$};
\node (S)  [ right of=V, node distance=3.5cm] {$S_{i}$};
\node (A)  [ right of=S, node distance=1cm] {$A_{i}$};
\node (O)  [ right of=A, node distance=4cm] {$O_{i}$};
 \node (Y)  [above right of=A,node distance=10mm]{$Y$};
  \node (N)  [below right of=A, node distance=10mm] {$N$};
 \node (YY)  [right of=Y, node distance=2.5cm] {$Y$};
 \node (NN)  [ right of=N, node distance=2.5cm] {$N$};
  
\draw[->] (G) to node[above]{$p$} (H);
\draw[->] (G) to node[sloped, anchor=center,above, pos=.3]{$1-p$} (L);
\draw[->] (B) to node[sloped, anchor=center,below, pos=.2]{$1-p$} (H);
\draw[->] (B) to node[below]{$p$} (L);

\draw[->] (Y) to node[above]{$1-\beta$} (YY);
\draw[->] (Y) to node[above, pos=.3]{$\beta$} (NN);
\draw[->] (N) to node[sloped, anchor=center,below, pos=.2]{$\epsilon$} (YY);
\draw[->] (N) to node[below]{$1-\epsilon$} (NN);

\end{tikzpicture}
\caption{The Binary Symmetric Chnnel though which agents receive signals}
\label{fig1}
\end{figure}

 Consider the first agent, as there are no past observations, his optimal action is to follow $S_1$, i.e. to buy the product only if he receives a H(High) signal. Now, for the $n^{th}$ agent’s decision, let the prior observation history be denoted by $H_{n-1}=\{{O_{1},O_{2},\ldots,O_{n-1}}\}$. Each agent then takes the Bayesian optimal action given $H_{n-1}$ and $S_n$ which is defined as follows.

Let $\gamma_{n}=\mathbb{P}(G|S_{n},H_{n-1})$ denotes the posterior probability for the true value of item is good, $V=G$. 

Then the Bayesian optimal action taken by agent $n$ is defined as:
\begin{equation}
A_{n}=\begin{cases}
Y, & \text{if} \ \gamma_{n} >1/2\\
N, &\text{if} \ \gamma_{n} <1/2\\
follows\, S_{n},  &\text{if} \ \gamma_{n} =1/2\\
\end{cases}
\label{XX}
\end{equation}

Hence, from now on, we assume that each agent $n$ takes action $A_{n}$ based on his private signal $S_{n}$ and noisy observations ${O_{1},O_{2},\ldots,O_{n-1}}$ of all past agents' actions ${A_{1},A_{2},\ldots,A_{n-1}}$. In the next chapter, we will discuss several scenarios based on this model.

\section{Information Cascade}
$Definition$ 1. The public likelihood ratio is defined as $l_{n}(H_{n})=\mathbb{P}(H_{n}|B)/\mathbb{P}(H_{n}|G)$. The private likelihood ratio of agent $n$, $\delta_{n}$, is defined as $\delta_{n}(S_{n})=\mathbb{P}(S_{n}|B)/\mathbb{P}(S_{n}|G)$.

Using the above definition, we have $\gamma_{n} = 1/(1+\delta_{n}l_{n-1})$. Combine with \ref{XX}, in a decision-making process, a Y is the choice whenever $\delta_{n}l_{n-1}<1$. We name this case as information cascade, which is defined as follows:

$Definition$ 2. An information cascade is said to occur when a new agent takes a fixed action regardless of his private signal.

Now, if agent $n$ happens to be in an information cascade, then his optimal decision as described in \eqref{XX} implies that $\gamma_{n}>1/2(<1/2)$ for each $S_{n}\in\{H,L\}$ and the $n^{th}$ agent is said to be in a Y(N) cascade. The probability is when $\gamma \geq 1/2$ for $S_{n}= High$ and $\gamma \leq 1/2$ for $S_{n}= Low$ in which case, $A_{n}$ follows $S_{n}$.

$Lemma$ 1: Agent $n$ cascades to the action Y(N) when $l_{n-1}<\frac{1-p}{p}$($l_{n-1}>\frac{1-p}{p}$) and otherwise follows its private signal $S_{n}$.

And once a cascade happens, the public likelihood ratio $l_{n}$ would stay as the same because the observation $O_{n}$ does not bring any new information about the true value $V$ to the successors. Therefore, $l_{n+i}=l_{n-1}$ for all $i=0,1,2,\dots$, which is the following property.

$Property$ 1: Once a cascade happens, it stays forever.

Now, considering that agent $n$ is not in a cascade, Property 1 implies that $A_{i}$ follow $S_{i}$ for all $i=1,2,\dots,n-1$. Thus, the public likelihood ratio for the history $H_{n}$ can be expressed as $l_{n}=(\frac{1-p}{p})^{h_{n}}$, where $h_{n}=\eta_{Y}n_{Y}-\eta_{N}n_{N}$ is the weighted difference between the number of Y's($k_{Y}$) and the number of N's ($k_{N}$) in $H_{n}$. The weights $\eta_{Y}=\text{log}\frac{a}{1-b}/\text{log}\frac{p}{1-p}$ and $\eta_{Y}=\text{log}\frac{b}{1-a}/\text{log}\frac{p}{1-p}$ where for all $i=1,2,\dots,n-1$.

\begin{equation}
a=\mathbb{P}(O_{i}=Y|V=G) \ \text{and} \  b=\mathbb{P}(O_{i}=N|V=B).
\label{XXX}
\end{equation}
Denote the probabilities when $O_{i}$ follows the true value V given that $A_{i}$ follows $S_{i}$. 
From Figure \ref{fig1}, we have
\begin{equation}
a=p(1-\beta)+\epsilon(1-p) \ \text{and} \  b=p(1-\epsilon)+\beta(1-p)
\label{XXXX}
\end{equation}
Thus, until a cascade occurs, the likelihood ratio $l_{n}$ depends only on the number of Y's and N's present in the history through $h_{n}$ which thereby serves as a sufficient statistic of the information contained in the past observations, which give us the following property.

$Property$ 2: agents will follow their own private signal until cascade occurs.
Now, substituting the above obtained expression for $l_{n}$ in Lemma 1, we see that agent $n$ cascades to a Y (N) only when $h_{n} > 1$ ($h_{n} < -1$). Until then, the process $\{h_{n}\}$ updates as per the rule : information cascade happens when $h_{n}<-1$ or $h_{n}>1$, and starting at 0, the update rule for $h_{n}$ is:
\begin{equation}
h_{n}=\begin{cases}
h_{n-1}+\eta_{Y}, & \text{if} \ O_{n} =Y\\
h_{n-1}-\eta_{N}, &\text{if} \ O_{n} =N\\
\end{cases}
\label{XXXXX}
\end{equation}
and thereafter stops updating (Property 1).

\section{Markovian Analysis of Cascades}
It follows from last section that the random process $\{h_{n}\}$ is a countable state space DTMC that starts at $0$ and takes values in [-1,1] until it ends up in one of the two absorption states: the left wall $(-\infty, -1)$ and the right wall $(1,\infty)$ that correspond to a N cascade and Y cascade respectively. Each state represents the agent’s observation history. Moreover, until a cascade occurs, the update rule for $\{h_{n}\}$ in (3.3) describes it as a random walk that either takes a forward (rightward) step by $\eta_{Y}$ with probability $p_f$ or takes  a backward (leftward) step by $\eta_{N}$ with probability $1-p_f$. Here,  $p_f := \mathbb{P}(O_{n}=Y|V) $ refers to the "forward jump probability", i.e, the probability of a Y observation given the true state $V$. From \eqref{XX} we have
\begin{equation}
p_{f}=a \ \text{for} \ V=G \ \text{and} \  p_{f}=1-b \ \text{for} \ V=B.
\label{Y}
\end{equation}

\begin{figure}[h]
\begin{center}
\begin{tikzpicture}[scale=1]

\coordinate (P) at (0.5, 0);
\coordinate (A) at (1, 0);
\coordinate (B) at (2, 0);
\coordinate (C) at (3, 0);
\coordinate (D) at (4, 0);
\coordinate (E) at (5, 0);
\coordinate (F) at (6, 0);
\coordinate (G) at (7, 0);
\coordinate (Q) at (8, 0);

\draw[->] (P)--(Q);
\draw[-] (D) to [bend right] node[above]{$1-p_{f}$} (B);
\draw[-] (D) to [bend left] node[above]{$p_{f}$} (E);

\draw (A) node [below]{-1}  -- ++(0, 3pt) ;
\draw (B)  node [below]{$-\eta_{N}$}-- ++(0, 3pt);

\draw (D) node [below]{0}-- ++(0, 3pt) ;
\draw (E)  node [below]{$\eta_{Y}$} -- ++(0, 3pt) ;
\draw (G) node [below]{1}  -- ++(0, 3pt) ;

\end{tikzpicture}
\end{center}
\caption{Transition Diagram of the Random Walk}
\label{fig2} 
\end{figure}
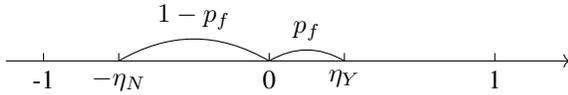
Figure \ref{fig2} depicts the random walk.

In \cite{label4}, the noise is unbiased towards each of the possible actions, which leads to $\eta_{Y}=\eta_{N}$ and finite states on Markov chain. Furthermore, in \cite{label5}, the noise is biased only towards to one preferred action that represented an uncountable stat-space Markov chain. Note that in our case, we mainly consider the non-integer values of $1/\eta_{Y}$ and $1/\eta_{N}$, which means the $h_{n}$ needs to take uncountable steps to get one of absorbing states. Because when $\eta_{Y}$ and $\eta_{N}$ satisfies $1/\eta_{Y}$ and $1/\eta_{N}=r$ for some $r=1,2,3,\dots$ the random walk is a sample Markov Chain with finite state and having two absorbing states represented Y and N cascade.

For the random walk $\{h_n\}$, given the true value $V \in \{G, B\}$, let $P_{Y}^{V}$ denote the probability of a Y cascade. The N cascade probability denoted by $P_{N}^{V}$ is simply $1-P_{Y}^{V}$, as a cascade occurs almost surely.

\subsection{Y cascade probability}
In this subsection, we plot the probability of a wrong cascade under $V=B$, i.e., $P_{Y}^{B}$ for $\epsilon$ varying from 0 to $1-\beta$ for a fixed $\beta$. Refer to Fig. \ref{fig1} which compares two plots of $P_{Y}^{B}$ for values: $\beta = 0$ and $\beta =0.5$. Recall that $\beta = 0$ corresponds to the single fake agent-type scenario of \cite{label5}. Compared to one type of fake agent scenario, the occurrence of two types of fake agents shows that a certain decrease of wrong cascade. When we enroll the second type of fake agents, they can be treated as a compensation to the negative effect caused by the first type of fake agents. Counter to basic knowledge, the probability of Y cascade is not a monotonically increasing function with respect of $\epsilon$. There is a significant decrease at some certain $\epsilon$ value. This character will be discussed in the next section. Further, from the point of view of the fake agents, they would not benefit from their portion increasing as $\epsilon$ continue increasing. That is because as the probability of an agent being fake tends to 1, the information included in a Y observation is negligible. There is less and less information being considered from an observed Y.

\begin{figure}[h]
\centering
\includegraphics[height=8cm,width=9cm]{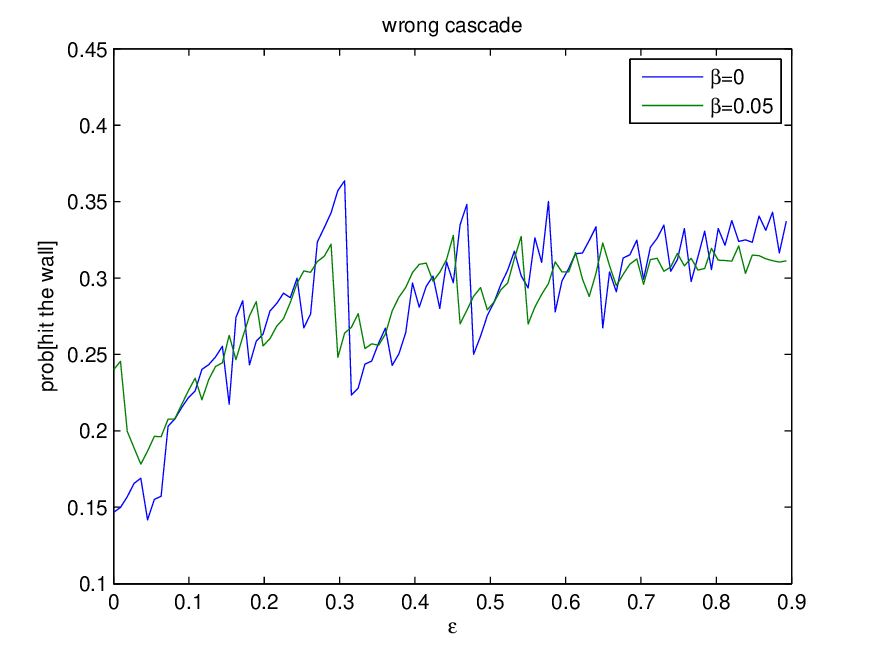}
\caption{Probability of Y Cascade as a function of $\epsilon$ for V=B, p=0.7 and $\beta=0$, $\beta=0.05$}
\label{3}
\end{figure}

\begin{figure}[h]
\centering    
\subfigure[$\beta=0.1$] {
\label{fig:a}     
\includegraphics[width=0.4\columnwidth]{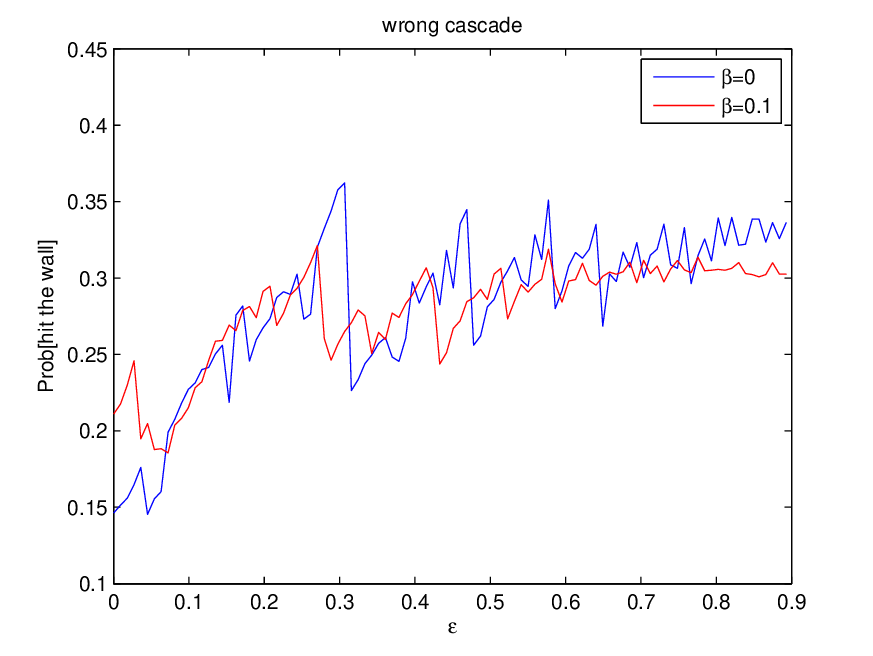}  
}      
\subfigure[$\beta=0.2$] { 
\label{fig:b}     
\includegraphics[width=0.4\columnwidth]{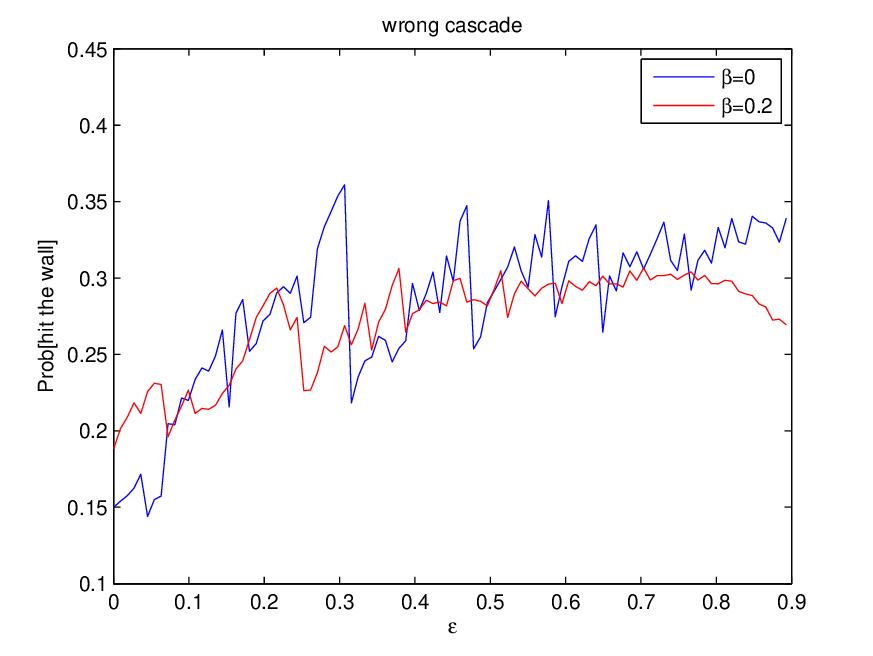}     
}   
\caption{ Probability of Y Cascade as a function of $\epsilon$ for V=B and p=0.7 with different $\beta$ }     
\label{fig4}     
\end{figure}

Next, we compare $P_{Y}^{B}$ for $\beta =0$ with the plots for $P_{Y}^{B}$ where $\beta =0.1, 0.2$ in Figure. \ref{fig4} respectively. We observe a general trend over most values of $\epsilon$ that an increase in the fraction of N type fake agents $\beta$ leads to a reduction in the wrong cascade prob. under $V=B$.

As the value of $\beta$ increase, the presence of second type of fake agents reduces the probability of wrong cascade. The fractional decrease at some certain $\epsilon$ shrinks as well. Figure. \ref{fig4} shows that the wrong cascade curve becomes more and more smooth, which we will discuss in next section.

\subsection{Error Thresholds}
In\cite{label5}, the author discussed the fractional decrease at some certain $\epsilon$ thresholds. At first when there is no fake agent, the sequence needs to observe one action two times to have a cascade. As the portion of fake agents increase, the number of consecutive Y needed to start a cascade increases because each observation provides less information. It can be proved in the following lemma.

$Lemma$ 2: Let $\alpha=p/(1-p)$. For $r=0,1,2,\dots$ define the sequence of thresholds $\{\epsilon_{r}\}^{\infty}_{r=1}$. By using Bayes' rule, the denotes the posterior probability $\gamma_{n}$ is given as:

\begin{equation}
\gamma_{n}=\frac{(1-p)a^{r}}{(1-p)a^{r}+p(1-b)^{r}}
\label{YYY}
\end{equation}
Agent $n$ does not have a cascade if $\gamma_{n}\leq\frac{1}{2}$, i.e. $(1-p)a^{r}\leq p(1-a)^{r}$. The $r^{th}$ threshold $\epsilon_{r}$ is given as:

\begin{equation}
\epsilon_{r}=(1-\beta)\frac{\alpha-\alpha^{\frac{1}{r}}}{\alpha^{\frac{1}{r}+1}-1}
\label{YY}
\end{equation}

\begin{figure}[h]
\centering
\includegraphics[height=8cm,width=9cm]{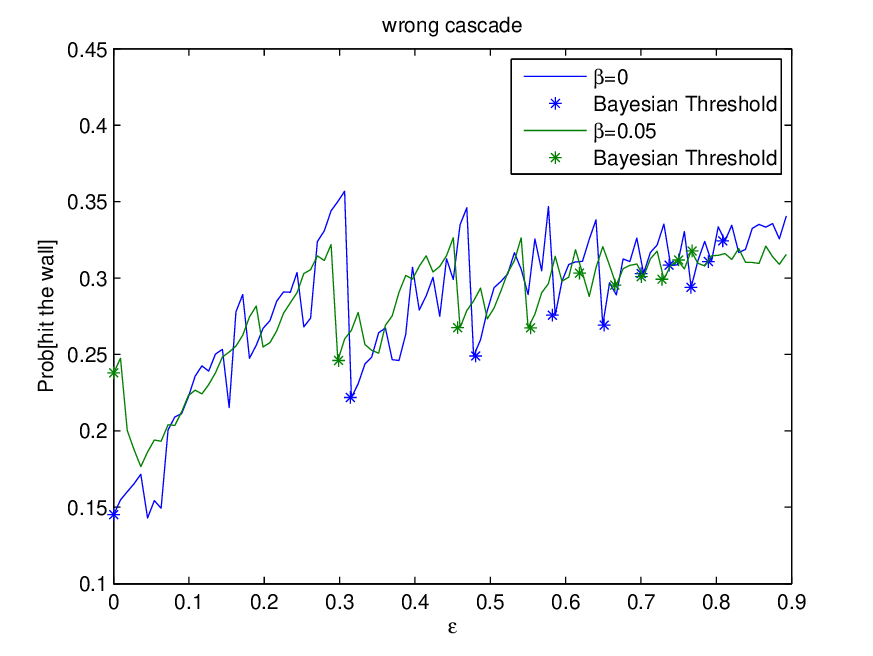}
\caption{Bayesian Thresholds}
\label{fig5}
\end{figure}

\begin{figure} \centering    
\subfigure[$\beta=0.1$] {
 \label{fig:a}     
\includegraphics[width=0.4\columnwidth]{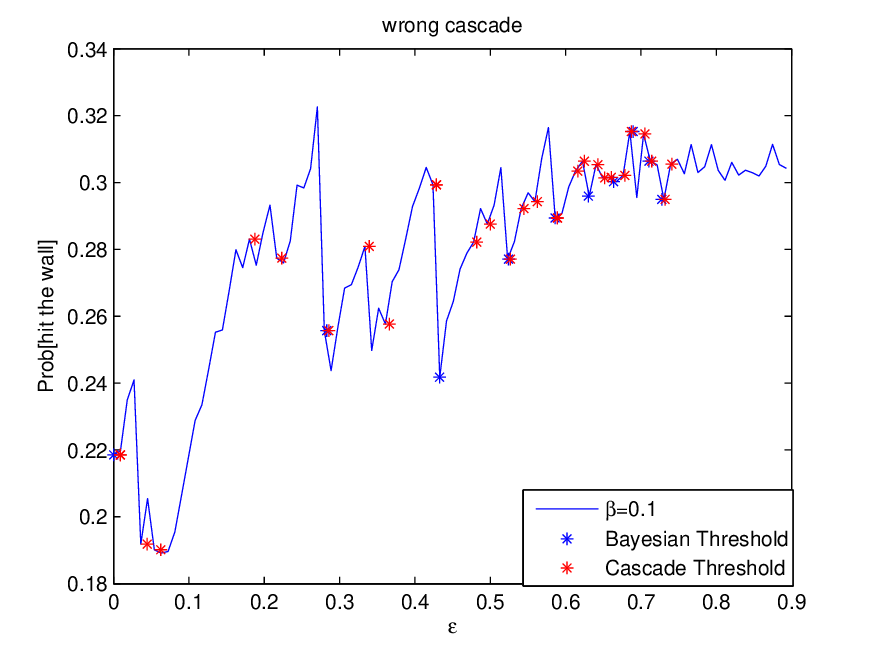}  
}     
\subfigure[$\beta=0.2$] { 
\label{fig:b}     
\includegraphics[width=0.4\columnwidth]{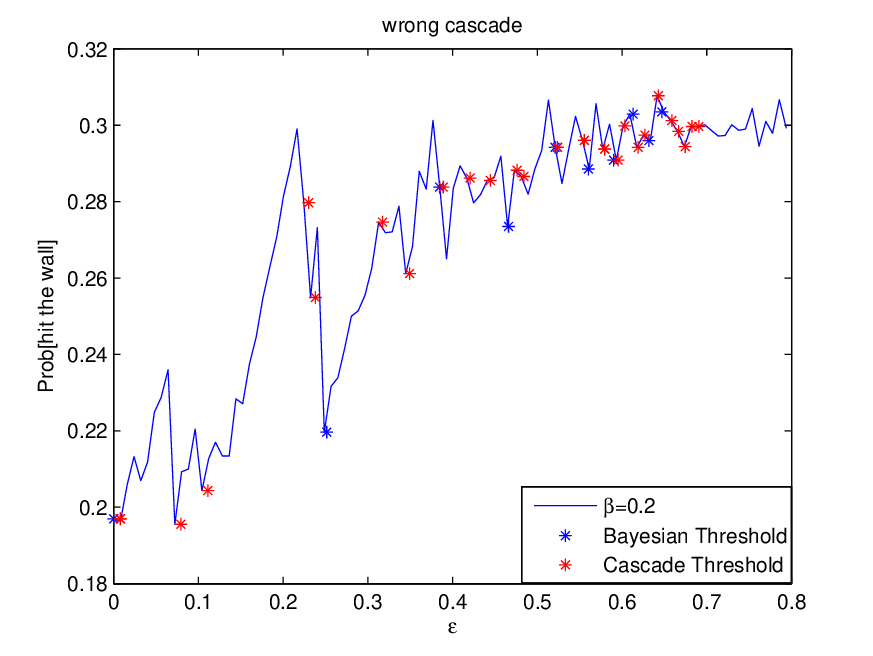}     
}    
\caption{ Bayesian Thresholds and Cascade Thresholds }     
\label{fig6}     
\end{figure}

We observe in Figure. \ref{fig5} that the relatively larger drops in $P_{Y}^{B}$ ($\epsilon$) (marked by *) occur exactly at the threshold points $\{\epsilon_{r} \}_{r=1}^{\infty}$. Here, a slight increase in $\epsilon$ beyond $\epsilon_{r}$ causes a significant decrease in the probability of a Y cascade. An intuitive reasoning for this reduction the different number of consecutive Ys are needed to start a cascade when there is only one N in the sequence. That leads to we wonder what would happen if the sequence had more than one N and how many consecutive Ys are needed. Following that idea, the cascade thresholds are defined as calculating the probability of wrong cascade based on multiple Ns in the sequence. As shows in the Figure. \ref{fig6}, Bayesian thresholds, second and third cascade thresholds capture most of discontinuities in the curve.

\subsection{Tree and Sequence Structure}
In this subsection, we outline two approaches for estimating the wrong cascade probability. A tree structure is proposed to estimate $P_{Y}^{B}$ and a sequence structure to discuss the lower bound.

\subsubsection{Tree Structure}

\begin{figure}[h]
\begin{center}
\begin{tikzpicture}[scale=1]

\coordinate (P) at (0.5, 0);
\coordinate (A) at (1, 0);
\coordinate (B) at (2, 0);
\coordinate (C) at (3, 0);
\coordinate (D) at (4, 0);
\coordinate (E) at (5, 0);
\coordinate (F) at (6, 0);
\coordinate (G) at (7, 0);
\coordinate (Q) at (8, 0);

\draw[->] (P)--(Q) node [below right]{$x$};

\draw (A) node [below]{-1}  -- ++(0, 3pt) ;
\draw (C)  node [below]{$\eta_{N}-1$}-- ++(0, 3pt);

\draw (D) node [below]{0}  node[above] {$comfort$ $  $  $zone$}-- ++(0, 3pt) ;
\draw (E)  node [below]{$1-\eta_{N}$} -- ++(0, 3pt) ;
\draw (G) node [below]{1}  -- ++(0, 3pt) ;

\end{tikzpicture}
\end{center}
\caption{Comfort Zone}
\label{fig7} 
\end{figure}
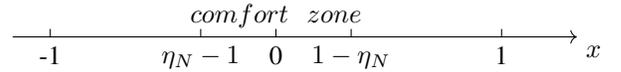

Let us consider a way to estimate the probability of a Y cascade. Because our focus is non-integer values of $1/\eta$ resulting in uncountable values of the number of needed consecutive Ys. Besides, due to two types of fake agents, sequence would be allowed multiple Ns without hitting the left side wall. It leads to infinite long sequence. To have an approximation of our model, we proposed a tree structure to describe our model. We firstly defined the comfort zone for every sequence with different $\eta_{N}$. The comfort zone is given as:
$[\eta_{N}-1,1-\eta_{N}]$. 
Then, we divide the whole region [-1,1] into several sub-regions, such as $[-1,\eta_{N}-1]$, $[\eta_{N}-1,0]$, $[0,1-\eta_{N}]$ and $[1-\eta_{N},1]$. The following iterative process depicted in Figure. \ref{fig7} describes all possible sequences that can lead to a Y cascade.

The iteration process shows in Figure. \ref{fig7}, $r_{0}=\lceil \frac{1}{\eta_{Y}} \rceil$, $r_{1}=\lceil \frac{\eta_{N}-1}{\eta_{Y}} \rceil$, $r_{2}=\lceil \frac{1-\eta_{N}}{\eta_{Y}} \rceil$. Let M denotes the number of iterations. The sequence will be starting at 0, moving forward with probability $P_{f}$ and backward with probability $1-P_{f}$. For every time the sequence come back to the comfort zone, a new iteration will begin until the sequence hit one of walls. Note that the  comfort zone is not the same as before for a new iteration process. Because it is unlikely the sequence would land exactly at 0 when it comes back to the comfort zone. Hence, for the accuracy of our approximation, a new strating point and comfort zone will be defined before the next iteration process. 

\begin{figure}[h]
\centering
\includegraphics[height=6cm,width=8cm]{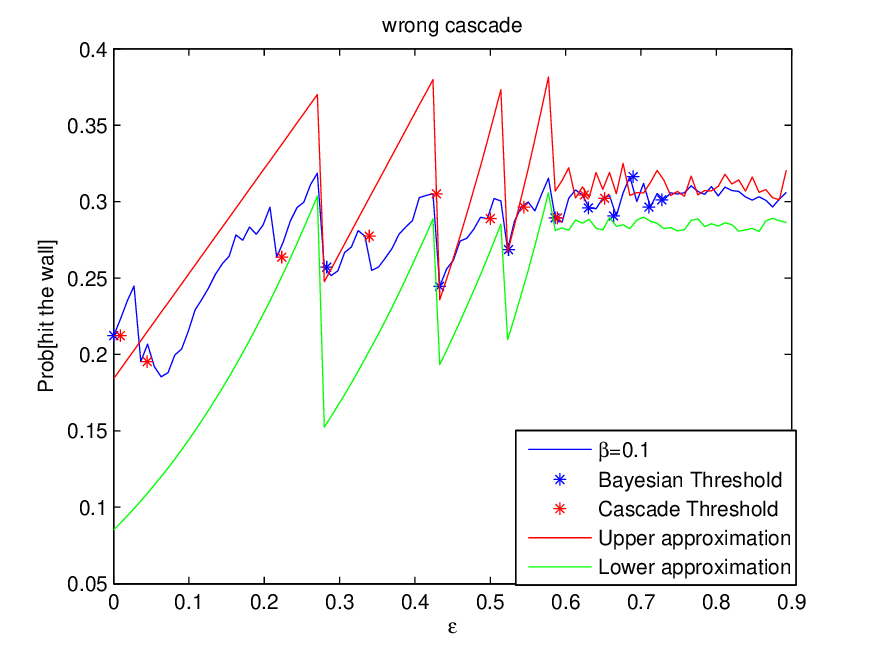}
\caption{Upper and Lower Approximation with Thresholds}
\label{fig8}
\end{figure}

In Figure. \ref{fig8}, we draw the upper and lower approximation of our model based on this iteration rule. The plot uses M=10 which gives a close approximation. Our approximation pefectly captures many discontinuities at Bayesian and Cascade Thresholds. Note that the simulation is outside of our approximation when $\epsilon$ is small. It did not expect a fractional decrease there.

\subsubsection{Sequence Structure}
In order to have an accurate lower bound for the probability of wrong cascade, we proposed another sequence structure to represent all possible sequences as Figure.\ref{fig10} shown. In this case, we divide positive region[0,1] into (K+2) sub-regions based on $\eta_{Y}$. Once the sequence land into the $(K+1)^{th}$ stage, it only needs one Y to have a Y cascade. The way we count all possible sequences is starting at 0 and moving forward stage by stage.

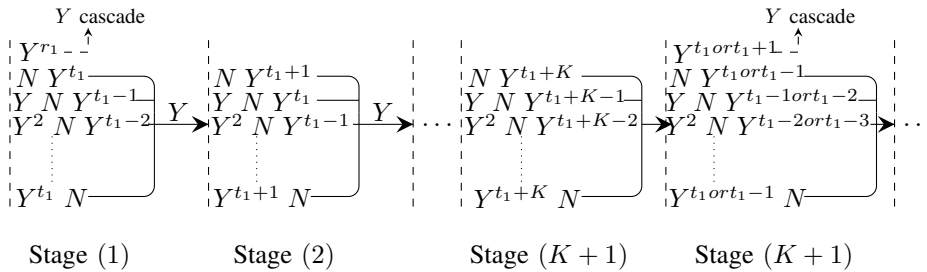
\begin{figure*}[hb]
\centering
\begin{tikzpicture}[scale=1.6]
\begin{scope}[shift={(-4,0)}]
\draw [dashed, decoration={markings,mark=at position 1 with {\arrow[scale=1,>=stealth]{>}}},postaction={decorate}] (0.1,0) -- (0.3,0) -- (0.3,0.2); \node at (0.4,0.3) {\scalebox{0.8}{$Y$ cascade}};
\draw [rounded corners, decoration={markings,mark=at position 1 with {\arrow[scale=2,>=stealth]{>}}},postaction={decorate}] (0.8,-0.6) -- (1.3,-0.6);
\node at (1.05,-0.5) {$Y$}; 
\node at (-0.1,0) {$Y^{r_1}$}; 
\node at (0,-0.2) {$N \; Y^{t_1}$};  \draw [rounded corners] (0.3,-0.2) -- (0.85,-0.2)--(0.85,-0.5);
\node at (0.2,-0.4) {$Y \; N \; Y^{t_1-1}$}; \draw [rounded corners] (0.7,-0.4) -- (0.85,-0.4);
\node at (0.25,-0.6) {$Y^2 \; N \; Y^{t_1-2}$}; 
\draw [dotted] (0,-0.7) -- (0,-1.1);
\node at (0,-1.2) {$Y^{t_1} \; N$}; \draw [rounded corners] (0.3,-1.2) -- (0.85,-1.2)--(0.85,-0.5);
\draw [dashed] (-0.35,0.1) -- (-0.35,-1.3);
\draw [dashed] (1.3,0.1) -- (1.3,-1.3);
\node at (0.23,-1.7) {Stage $(1)$};
\end{scope}
\begin{scope}[shift={(-2.3,0)}]
\draw [rounded corners, decoration={markings,mark=at position 1 with {\arrow[scale=2,>=stealth]{>}}},postaction={decorate}] (0.8,-0.6) -- (1.3,-0.6);
\node at (1.05,-0.5) {$Y$}; 
\node at (0.05,-0.2) {$N \; Y^{t_1+1}$};  \draw [rounded corners] (0.5,-0.2) -- (0.85,-0.2)--(0.85,-0.5);
\node at (0.05,-0.4) {$Y \; N \; Y^{t_1}$}; \draw [rounded corners] (0.5,-0.4) -- (0.85,-0.4);
\node at (0.2,-0.6) {$Y^2 \; N \; Y^{t_1-1}$}; 
\draw [dotted] (0,-0.7) -- (0,-1.1);
\node at (0.05,-1.2) {$Y^{t_1+1} \; N$}; \draw [rounded corners] (0.45,-1.2) -- (0.85,-1.2)--(0.85,-0.5);
\draw [dashed] (1.3,0.1) -- (1.3,-1.3);
\node at (0.23,-1.7) {Stage $(2)$};
\node at (1.5,-0.6) {$\ldots$};
\end{scope}

\begin{scope}[shift={(-0.1,0)}]
\draw [rounded corners, decoration={markings,mark=at position 1 with {\arrow[scale=2,>=stealth]{>}}},postaction={decorate}] (1,-0.6) -- (1.25,-0.6);
\node at (0,-0.2) {$N \; Y^{t_1+K}$};  \draw [rounded corners] (0.5,-0.2) -- (1,-0.2)--(1,-0.5);
\node at (0.2,-0.4) {$Y \; N \; Y^{t_1+K-1}$}; \draw [rounded corners] (0.85,-0.4) -- (1,-0.4);
\node at (0.25,-0.6) {$Y^2 \; N \; Y^{t_1+K-2}$}; 
\draw [dotted] (0,-0.7) -- (0,-1.1);
\node at (0.05,-1.2) {$Y^{t_1+K} \; N$}; \draw [rounded corners] (0.5,-1.2) -- (1,-1.2)--(1,-0.5);
\draw [dashed] (-0.5,0.1) -- (-0.5,-1.3);
\draw [dashed] (1.2,0.1) -- (1.2,-1.3);
\node at (0.23,-1.7) {Stage $(K+1)$};
\end{scope}
\begin{scope}[shift={(1.5,0)}]
\draw [dashed, decoration={markings,mark=at position 1 with {\arrow[scale=1,>=stealth]{>}}},postaction={decorate}] (0.5,0) -- (0.7,0) -- (0.7,0.2); \node at (0.8,0.3) {\scalebox{0.8}{$Y$ cascade}};
\draw [rounded corners, decoration={markings,mark=at position 1 with {\arrow[scale=2,>=stealth]{>}}},postaction={decorate}] (1.3,-0.6) -- (1.5,-0.6);
\node at (0.1,0) {$Y^{t_1 or t_1+1}$};
\node at (0.2,-0.2) {$N \; Y^{t_1 or t_1-1}$};  \draw [rounded corners] (0.75,-0.2) -- (1.35,-0.2)--(1.35,-0.5);
\node at (0.4,-0.4) {$Y \; N \; Y^{t_1-1 or t_1-2}$}; \draw [rounded corners] (1.2,-0.4) -- (1.35,-0.4);
\node at (0.45,-0.6) {$Y^2 \; N \; Y^{t_1-2 or t_1-3}$}; 
\draw [dotted] (0,-0.7) -- (0,-1.1);
\node at (0.2,-1.2) {$Y^{t_1 or t_1-1} \; N$}; \draw [rounded corners] (0.75,-1.2) -- (1.35,-1.2)--(1.35,-0.5);
\draw [dashed] (1.5,0.1) -- (1.5,-1.3);
\node at (0.5,-1.7) {Stage $(K+1)$};
\node at (1.7,-0.6) {$\ldots$};
\end{scope}
\end{tikzpicture}
\caption{\small An enumeration of all possible sequences that would lead to a $Y$ cascade.}  
\label{iter_struct}
\end{figure*}

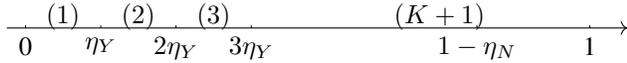
\begin{figure}[h]
\begin{center}
\begin{tikzpicture}[scale=0.5]

\coordinate (P) at (0.5, 0);
\coordinate (A) at (1, 0);
\coordinate (B) at (3, 0);
\coordinate (C) at (5, 0);
\coordinate (D) at (7, 0);
\coordinate (E) at (11, 0);
\coordinate (F) at (13, 0);
\coordinate (G) at (16, 0);
\coordinate (Q) at (17, 0);

\draw[->] (P)--(Q) ;

\draw (A) node [below]{0}  -- ++(0, 2pt) ;
\draw (B)  node [below]{$\eta_{Y}$}-- ++(0, 2pt);
\draw (C)  node [below]{$2\eta_{Y}$}-- ++(0, 2pt);
\draw (D) node [below]{$3\eta_{Y}$}  -- ++(0, 2pt) ;
\draw (F)  node [below]{ $1-\eta_{N}$} -- ++(0, 2pt) ;
\draw (G) node [below]{1}  -- ++(0, 2pt) ;

\node at (2,0.3) {$(1)$}(0, 1pt);
\node at (4,0.3) {$(2)$};
\node at (6,0.3) {$(3)$};
\node at (12,0.3) {$(K+1)$};
\end{tikzpicture}
\end{center}
\caption{Stage of Sequence Strcuture}
\label{fig10} 
\end{figure}

 We assume $\eta_{N}$ $>$ $\eta_{Y}$. The counting process starts at 0. Let $r_{1}=\lceil \frac{1}{\eta_{Y}}\rceil$, denotes the number of consecutive Y needed to hit the right wall starting at 0, $t_{1}=\lceil \frac{\eta_{N}}{\eta_{Y}}\rceil$, denotes the number of consecutive Y needed to compensate from a N to come back. The relationship between these two parameters is given as:
\begin{equation}
1-\eta_{N}=(K+1)\eta_{Y} \Rightarrow K+1=r_{1}-t_{1}
\end{equation}
Each stage includes all possible sequences landed in this region, including multiple Ns cases. We can calculate the whole stage probability of all the possible sequences at one time. After the counting process arrives $(K+1)^{th}$ state, the iteration begins. Here the $(K+1)^{th}$ stage is much similar as the comfort zone we talked about in the last section.

\begin{figure}[h!]
\centering
\includegraphics[height=5cm,width=7cm]{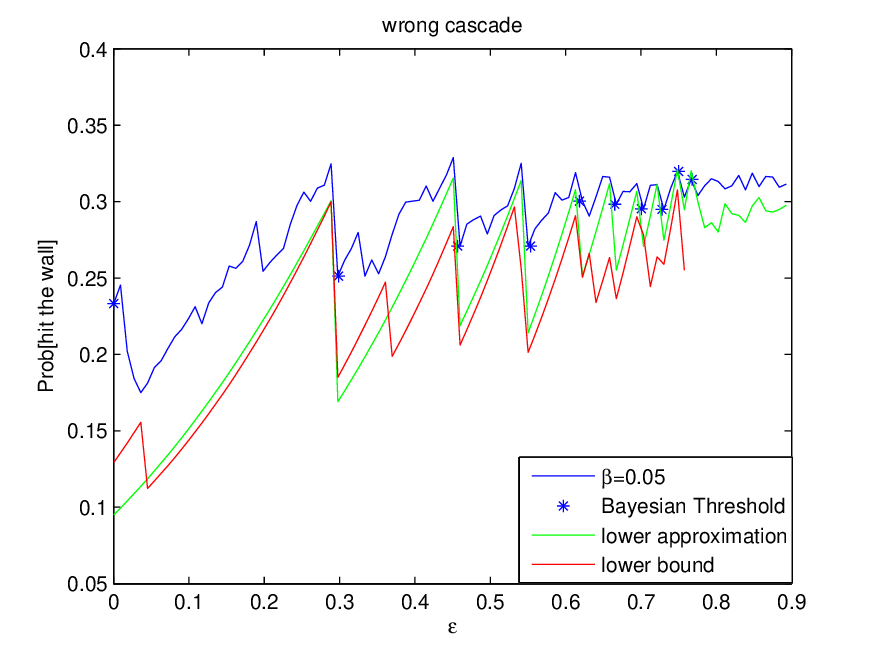}
\caption{Lower Bound and Lower Approximation with Bayesian Thresholds}
\label{fig11}
\end{figure}

Figure. \ref{fig11} shows both the lower bound and lower approximation with Bayesian Thresholds. The plot uses $M=10$ which gives an error of less than $10^{-4}$.

\section{Conclusions}
In this paper, We studied the effect of randomly arriving two types of fake agents that seek to influence the outcome of an information cascade. We concentrated on the impact of varying one type of fake agents on the probability of their preferred cascade. Using a Markov chain based analysis we determined two types of  thresholds, i.e. Bayesian and Cascade thresholds. In addition, we employed a tree structure and sequence structure to have an approximation and a lower bound of the probability of wrong cascade. In future work we plan on a non-Bayesian rationality network, and the limit when there is a small amount of fake agents.


\begin{thebibliography}{99}




\bibitem{label1}  S. Bikhchandani, D. Hirshleifer, and I. Welch, “A theory of fads, fashion, custom, and cultural change as informational cascades,” \textit{Journal of political Economy}, vol. 100, no. 5, pp. 992–1026, 1992.
\bibitem{label2} A. V. Banerjee, “A simple model of herd behavior,” \textit{The quarterly journal
of economics}, vol. 107, no. 3, pp. 797–817, 1992. 
\bibitem{label3} I. Welch, “Sequential sales, learning, and cascades,” \textit{The Journal of
finance}, vol. 47, no. 2, pp. 695–732, 1992.
\bibitem{label4} T. N. Le, V. G. Subramanian, and R. A. Berry, “Information cascades
with noise,” \textit{IEEE Transactions on Signal and Information Processing
over Networks}, vol. 3, no. 2, pp. 239–251, 2017.
\bibitem{label5} P. Poojary and R. Berry, "Observational Learning with Fake Agents," \textit{2020 IEEE International Symposium on Information Theory (ISIT)}, 2020, pp. 1373-1378.\bibitem{label11} P. Poojary and R. Berry, "Observational Learning with Fake Agents," \textit{2020 IEEE International Symposium on Information Theory (ISIT)}, 2020, pp. 1373-1378.
\bibitem{label6} D. Acemoglu, M. A. Dahleh, I. Lobel, and A. Ozdaglar, “Bayesian
learning in social networks,” \textit{The Review of Economic Studies}, vol. 78,
no. 4, pp. 1201–1236, 2011.
\bibitem{label7}  L. Smith and P. Sørensen, “Pathological outcomes of observational
learning,” \textit{Econometrica}, vol. 68, no. 2, pp. 371–398, 2000.
\bibitem{label8} D. Acemoglu, G. Como, F. Fagnani, and A. Ozdaglar, “Opinion fluctuations and disagreement in social networks,” \textit{Mathematics of Operations Research}, vol. 38, no. 1, pp. 1–27, 2013.
\bibitem{label9} L. Smith, P. Sorensen, "Pathological Outcomes of Observational Learning", \textit{Econometrica}, vol. 68, pp. 371-398, 200
\bibitem{label10}  Y. Wang, P. Djuric, "Social learning with Bayesian agents and random decision making", \textit{IEEE Trans Sig. Process}, vol. 63, no. 12, 2015.
\bibitem{label11} Y. Song, "Social Learning with Endogenous Network Formation", submitted to J. Econ. Theory, 2015.








\end{thebibliography}
\end{document}